# Numerical investigation of evapotranspiration processes in a forested watershed of Central Siberia

Laurent Orgogozo[1]
Anatoly S. Prokushkin[2]
Oleg S. Pokrovsky[1,3]
Christophe Grenier[4]
Michel Quintard[5,6]
Jérôme Viers[1]
Stéphane Audry[1]

[1]*GET (Géosciences Environnement Toulouse), UMR 5563 CNRS / UR 234 IRD / UPS, Observatoire Midi-Pyrénées, Université de Toulouse, 14 avenue Édouard Belin, 31400 Toulouse, France, contact: laurent.orgogozo@get.omp.eu*
[2]*V.N. Sukachev Institute of Forest, Siberian Branch, Russian Academy of Sciences, Akademgorodok 50/28, Krasnoyarsk, Russia*
[3]*BIO-GEO-CLIM Laboratory, Tomsk State University, Lenina 35, Tomsk, Russia*
[4]*LSCE/IPSL (Laboratoire des Sciences du Climat et de l'Environnement), UMR 8212 CNRS-CEA-UVSQ, CEA - Orme des Merisiers, 91191 Gif-sur-Yvette Cedex, France*
[5]*Université de Toulouse, INPT, UPS, IMFT (Institut de Mécanique des Fluides de Toulouse), Allée Camille Soula, F-31400 Toulouse, France*
[6]*CNRS, IMFT: F-31400 Toulouse, France*

## Abstract

Evapotranspiration has a major control on continental surfaces dynamics in forested boreal environments. For example, it has strong impacts on active layer thickness, mainly because of its effect on water content within the upper layers of the soil column. These are complex processes, depending not only on climate forcings (atmospheric water demand) but also on physical, geo-pedological and biological properties of the considered areas. Here we propose a numerical investigation of evapotranspiration processes in the active layer of slopes of a forested watershed in Central Siberia. The effect on actual evapotranspiration of the spatial contrasts in terms of exposure, root layer thickness and tree stand density within the watershed are simulated using a recently developed high performance computing cryohydrogeological model, permaFoam.

## Introduction

Evapotranspiration is one of the main driving factors of landscape variability in taïga environment, as it can be seen for example through the contrast between north and south aspected slopes in the region of Tura (see Figure 1). Modifications of evapotranspiration regimes in boreal areas, either in relation with climate changes or not, may have strong impacts on continental boreal surfaces dynamics, for instance on water fluxes (*e.g.* Duan et al., 2017). Thus in order to develop predictive modeling of the evolution of boreal continental surfaces under climate change, an accurate quantification of the evapotranspiration processes is required.

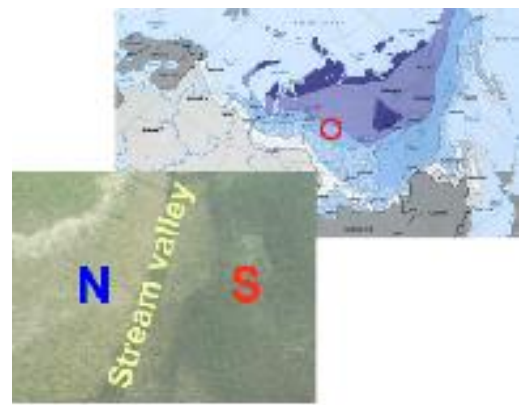

Figure 1: Visible contrast between north and south aspected slopes in a larch forest environment of Central Siberia.

The goal of this work is thus to develop a relevant quantification methodology for actual evapotranspiration in permafrost affected forested areas.

## Modelling evapotranspiration in the Kulingdakan watershed

In order to study the relative importance of different features of the forested boreal surfaces on evapotranspiration, the thermo-hydrological dynamics of the Kulingdakan experimental watershed (*e.g.* Prokushkin *et al.*, 2011, Viers *et al.*, 2015) have been simulated under current climatic conditions using permaFoam, a high performance computing (HPC) cryohydrogeology modeling tool (Orgogozo et al., 2015). The need of HPC technics in cryohydrogeology modeling is acknowledged (*e.g.* Painter et al., 2013), and that is the reason why we choose to develop permaFoam in the framework of the reference open source environment for computational fluid dynamics OpenFOAM® (*e.g.* Weller et al., 1998, www.openfoam.com). Based on estimated potential evapotranspirations (*i.e.* climate observations), the computation of the evapotranspirative uptakes in the active layers of the Kulingdakan watershed are performed in a temporally and spatially distributed way, taking into account both water availability and thermal status in the soils (modified from Orgogozo, 2015). Such a methodology allows the testing of the sensitivity of the actual evapotranspiration with respect to various properties of the slopes (*e.g.* exposure, root layer thickness, density of tree stand). The comparisons between observations and modeling results in terms of soil thermal regimes and water fluxes variability in the watershed allow discussing the relevance of each of the derived parameterizations of the evapotranspiration rates in the active layer.